\documentclass[12pt]{article}

\usepackage{amsfonts}
\usepackage{amssymb}
\usepackage{amsmath}
\usepackage{amsthm}
\usepackage{indentfirst}
\usepackage{epsfig}
\usepackage{graphicx}
\usepackage{mathrsfs}
\usepackage{natbib}
\usepackage{bm, url}
\usepackage{epic}
\usepackage{curves}

\newtheorem{theorem}{Theorem}[section]
\newtheorem{assw}{Assumption}

\newtheorem{corol}[theorem]{Corollary}

\newtheorem{exam}{Example}[section]
\newenvironment{example}{\begin{exam} \rm }{\hfill $\triangleleft$ \end{exam}}
\newtheorem{rema}{Remark}[section]

\newtheorem{defin}{Definition}[section]

\newtheorem{ass}[assw]{Assumption}

\makeatletter

\numberwithin{equation}{section} %
\numberwithin{figure}{section}

\begin{document}

\title{The minority game: An economics perspective}

\author{Willemien Kets\thanks{Address: Tilburg University, P.O. Box 90153, 5000 LE
Tilburg, The Netherlands. E-mail: w.kets@uvt.nl. Tel:
+31-13-4662478. Fax: +31-13-4663280. I am indebted to Ginestra
Bianconi, George Ehrhardt, Doyne Farmer, Matteo Marsili, Esteban
Moro, Jan Potters, Dolf Talman, and Mark Voorneveld for inspiring
discussions and helpful comments and suggestions. In addition, I
would like to thank Esteban Moro for his kind permission for
reproducing some of the figures from \citet{Moro2003}. All
remaining errors are of course my own.}}

\date{June 27, 2007}

\maketitle

\thispagestyle{empty}

\begin{abstract}
This paper gives a critical account of the minority game
literature. The minority game is a simple congestion game: players
need to choose between two options, and those who have selected
the option chosen by the minority win. The learning model proposed
in this literature seems to differ markedly from the learning
models commonly used in economics. We relate the learning model
from the minority game literature to standard game-theoretic
learning models, and show that in fact it shares many features
with these models. However, the predictions of the learning model
differ considerably from the predictions of most other learning
models. We discuss the main predictions of the learning model
proposed in the minority game literature, and compare these to
experimental findings on congestion games.
\end{abstract}



\newpage

\section{Introduction}

Congestion games are ubiquitous in economics. In a congestion game
\citep{Rosenthal1973}, players use several facilities from a
common pool. The costs or benefits that a player derives from a
facility depends on the number of users of that facility. A
congestion game is therefore a natural game to model scarcity of
common resources. Examples of such systems include vehicular
traffic \citep{NagelRasmussenBarrett1997}, packet traffic in
networks \citep{HubermanLukose1997}, and ecologies of foraging
animals \citep{DeAngelisGross1992}. Similar coordination problems
are encountered in market entry games \citep{SeltenGuth1982}.

Congestion games are also interesting from a theoretical point of
view. In congestion games, players need to coordinate to
differentiate. This seems to be more difficult than coordinating
on the same action, as any commonality of expectations is broken
up. For instance, when commuters have to choose between two roads
$A$ and $B$ and all believe that the others will choose road $A$,
nobody will choose that road, invalidating beliefs. The sorting of
players predicted in the pure-strategy Nash equilibria of such
games violates the common belief that in symmetric games, all
rational players will evaluate the situation identically, and
hence, make the same choices in similar situations \citep[see][p.
73]{HarsanyiSelten1988}. Moreover, in congestion games, players
may obtain asymmetric payoffs in equilibrium which may complicate
attainment of equilibrium, as coordination cannot be achieved
through tacit coordination based on historical precedent
\citep[cf.][]{Meyer_et_al_1992}. Finally, congestion games often
have many equilibria, so that players also face the difficulty of
coordinating on the same equilibrium.

Nevertheless, the theory of learning in games provides sharp
predictions on players' behavior in congestion games. As
congestion games belong to the class of potential games
\citep{MondererShapley1996}, all results that have been derived
for potential games apply to the class of congestion
games.\footnote{See e.g. \citet{HofbauerHopkins2005},
\citet{HofbauerSandholm2002}, \citet{MondererShapley1996}, and
\citet{Sandholm2001, Sandholm2007}. \citet{KetsVoorneveld2007}
study the convergence of play under different learning processes
in the minority game.} Experimental evidence, however, is not
always in line with these predictions. Though several experimental
studies have shown that players are remarkably successful at
learning to coordinate in congestion games,\footnote{For instance,
interacting players rapidly achieve a ``magical'' degree of tacit
coordination in market entry games, which is accounted for on the
aggregate level by the Nash equilibrium solution
\citep{Kahneman1988, Rapoport1995, Sundali_et_al_1995,
ErevRapoport1998, Rapoport_et_al_1998, Rapoport_et_al_2000}. See
e.g. \citet{Meyer_et_al_1992} and \citet{Selten2002} for similar
results on related games.} regularities on the aggregate level
generally conceal non-equilibrium behavior at the individual
level. Even though aggregate play is close to the Nash
equilibrium, individual players generally do not play equilibrium
strategies.\footnote{See e.g. \citet{Meyer_et_al_1992},
\citet{ErevRapoport1998}, \citet{Selten2002},
\citet{BottazziDevetag2004}.} Moreover, providing players with
more information does not always lead to better
outcomes.\footnote{For instance, in their experiments on market
entry games, \citet{ErevRapoport1998} find that providing players
with information on other players' actions may actually lead to
\emph{lower} average payoffs.}

These experimental findings are hard to explain with standard
learning models. This paper discusses the literature on the
minority game, a simple congestion game based on the El Farol bar
problem of \citet{Arthur1994}. Players have to choose between two
alternatives. Only those who have chosen the minority side get a
positive payoff. The minority game literature proposes a learning
model that is able to account for many of the experimental
findings listed above. We relate this learning model to the
standard learning models in economics, and compare its predictions
to experimental results on congestion games. The contribution of
the current paper is that it relates the literature on the
minority game, which has been largely developed in physics, to the
literature on learning in game theory and to the literature in
experimental economics on congestion games.\footnote{We have no
intention of giving a comprehensive survey of the minority game
literature, as an enormous amount of work on the minority game has
been done. For an extensive collection of papers on the minority
game, see \url{http://www.unifr.ch/econophysics/minority/}. See
\citet{Moro2003, ChalletMarsiliZhang2004} or \citet{Coolen2005}
for an introduction to the field. Papers in economics on the
minority game include \citet{BottazziDevetag2007},
\citet{ChmuraPitz2004}, and \citet{RenaultScarlattiScarsini2005}.
\citet{Blonski1999} and \citet{KojimaTakahashi2004} study learning
in games very similar to the minority game.}

\medskip

\noindent The outline of this paper is as follows. In
Section~\ref{sec:game_theoretic}, we introduce the minority game
and discuss its equilibria. The learning model proposed in the
minority game literature is discussed in
Section~\ref{sec:mg_model}. In Section~\ref{sec:results}, we
discuss the main predictions from the learning model. These
predictions are compared to experimental results on congestion
games in Section~\ref{sec:experiments}. Section \ref{sec:
conclusions} concludes.

\section{The stage game} \label{sec:game_theoretic}

The minority game is a game in which an odd number of players have
to choose between two actions; for instance, players either go to
a bar or stay home, either buy or sell an asset, etcetera. Players
want to distinguish themselves from the crowd: their aim is to
take a different action than the majority of players.

Following the notation of Tercieux and Voorneveld (2005), we
denote the set of players by $\mathcal{N} = \{1, \dots, 2k + 1\}$,
with $k \in \mathbb{N}$. Each player $i \in \mathcal{N}$ has a set
of \emph{pure strategies} $A_i = \{-1,+1\}$: agents have to choose
between two options.  The set of \emph{mixed strategies} of player
$i$ is denoted by $\Delta(A_i)$. We denote a mixed strategy
profile by $\alpha \in \times_{i \in \mathcal{N}} \Delta(A_i)$,
and we use the standard notation $\alpha_{-i} \in \times_{j \in
\mathcal{N} \setminus \{i\}} \Delta(A_j)$ to denote a strategy
profile of players other than $i \in \mathcal{N}$. With each
action $a \in \{-1,+1\}$, a function
\begin{equation*}
f_a : \{1, \dots, 2k+1\} \to \mathbb{R}
\end{equation*}
can be associated which indicates for each $n \in \{1, \dots,
2k+1\}$ the payoffs to a player choosing $a$ when the total number
of players choosing $a$ equals $n$. The von Neumann-Morgenstern
utility function of a player is then given by
\begin{equation}\label{utility}
u_i(a) = f_{a_i}\left(|\{j \in \mathcal{N} : a_j = a_i \}|\right),
\end{equation}
where $a \in \times_{j \in \mathcal{N}} A_j$. Payoffs are extended
to mixed strategies in the usual way.

The function $f_a(\cdot), a \in \{-1,+1\}$ can have several forms.
It is commonly assumed that congestion is costly:
\begin{quote}
\textbf{[Mon]} \qquad $f_{-1}$ and $f_{+1}$ are strictly
decreasing functions,
\end{quote}
and that the congestion effect is the same across alternatives:
\begin{quote}
\textbf{[Sym]} \qquad $f_{-1} = f_{+1}$.
\end{quote}
A commonly used form is $f_{-1}(n) = f_{+1}(n) = 1$ if $n \in \{1,
\dots, k\}$ and 0 otherwise \citep{ChalletZhang1997}.
Alternatively, one could define payoffs in terms of the aggregate
action $\sum_{i \in \mathcal{N}} a_i$ for a given action profile
$a = (a_i)_{i \in \mathcal{N}}$, with $a_i \in \{-1,+1\}$ for all
$i$. Let $g$ be a function on $\mathbb{R}$ such that $g(-x)=-g(x)$
for all $x \in \mathbb{R}$ and $g(x)> 0$ for $x>0$. A player $i
\in \mathcal{N}$ is then assigned the payoff
\begin{equation}\label{eq:g_payoff}
u_i(a) = -a_i g\left(\sum_{j \in \mathcal{N}} a_j\right).
\end{equation}
In our notation:
\begin{equation*}
f_{-1}(n) = f_{+1}(n) = g\left(2(k-n)+1\right).
\end{equation*}
Common choices include
\begin{equation} \label{eq:g_function}
g(x) = x / (2k+1)
\end{equation}
and
$$
g(x) = \text{sign}(x).
$$
Most of the predictions of the learning model are not affected
qualitatively by the precise choice of payoff function, given that
it satisfies [Mon] and [Sym] \citep{LiVanDeemenSavit2000}. Notice
that the minority game is a congestion game \citep{Rosenthal1973}
and hence a finite exact potential game
\citep{MondererShapley1996}.

\medskip

\noindent To analyze the game's Nash equilibria, we introduce some
more notation. A player who uses a mixed strategy that puts
positive probability on both pure strategies is referred to as a
\emph{mixer}. A player that puts full probability mass on the
alternative $-1$ is called a \emph{$(-1)$-player}; similarly, a
player that puts full probability mass on the alternative $+1$ is
called a \emph{$(+1)$-player}.

The stage game has a large number of Nash equilibria.
\citet{TercieuxVoorneveld2005} show that a pure strategy profile
is a Nash equilibrium if and only if one of the alternatives $-1$
or $+1$ is chosen by exactly $k$ of the $2k+1$ players. Note that
these Nash equilibria are \emph{not} strict, as a player that is
in the majority is indifferent between sticking to his choice or
switching actions, as his deviation would shift the majority.
There are $2\binom{2k + 1}{k}$ of such asymmetric pure-strategy
Nash equilibria.

\citet{KetsVoorneveld2007} characterize the game's mixed-strategy
Nash equilibria. It can be shown that in any Nash equilibrium with
at least one mixer, all mixers use the same mixed strategy.
Moreover, player labels are irrelevant by [Sym] (if $\alpha$ is a
Nash equilibrium, so is every permutation of $\alpha$). Together,
these facts imply that a Nash equilibrium with at least one mixer
can be summarized by its \emph{type} $(\ell, r, \lambda)$, where
$\ell, r \in \{0, 1, \ldots, 2k+1\}$ denote the number of players
choosing pure strategy $-1$ or $+1$, respectively, and $\lambda
\in (0,1)$ the probability with which the remaining $z(\ell, r,
\lambda) := (2k + 1) - (\ell + r) > 0$ mixers choose $-1$. Let
$v_{-1}(\ell, r, \lambda)$ denote the expected payoff to a player
choosing $-1$; $v_{+1}(\ell, r, \lambda)$ is defined similarly.
Then, a strategy profile of type $(\ell, r, \lambda)$ is a Nash
equilibrium if and only if
\begin{equation}\label{eq:equil_cond}
v_{-1}(\ell + 1, r, \lambda) = v_{+1}(\ell, r+1, \lambda),
\end{equation}
i.e., the expected payoffs to a mixer of playing the pure strategy
$a = -1$ are equal to the expected payoffs of the pure strategy $a
= +1$. It can be shown that there exist Nash equilibria with
exactly one mixer. These equilibria are of type $(k,k, \lambda)$
with arbitrary $\lambda \in (0,1)$, i.e., the mixer uses an
arbitrary mixed strategy, whereas the remaining $2k$ players are
spread evenly over the two pure strategies. In addition, there are
Nash equilibria with more than one mixer. For $\ell, r \in \{0, 1,
\ldots, 2k+1\}$ such that $\ell + r \leq 2k - 1$, there is a Nash
equilibrium of type $(\ell, r, \lambda)$ if and only if
$\max\{\ell, r\} < k$. The corresponding probability $\lambda \in
(0,1)$ solves \eqref{eq:equil_cond}, and can be shown to be
unique. It follows from these results that there is a unique
symmetric mixed-strategy Nash equilibrium. In this equilibrium,
each player chooses each option with probability $\frac{1}{2}$.
The expected number of players choosing each option is then
$k+\frac{1}{2}$.

\section{Learning in the minority game}\label{sec:mg_model}

Players in the minority game face both a coordination problem and
an incentive problem. The coordination problem is not easy to
solve. As the equilibria in pure strategies cannot be
Pareto-ranked or ordered in terms of risk-dominance, no particular
pure-strategy Nash equilibrium can be singled out as being most
salient \citep{Schelling1960}. Hence, without pre-play
communication, players do not have enough information to implement
a pure-strategy Nash equilibrium
\citep[cf.][]{MenezesPitchford2006}. While players could use
common knowledge of rationality and symmetry to deduce and select
the symmetric mixed-strategy Nash equilibrium
\citep[cf.][]{Ochs1990,Meyer_et_al_1992}, this may raise an
incentive problem, as players can earn a higher payoff than in the
symmetric mixed-strategy Nash equilibrium if they manage to
outsmart the other players. Hence, players may try to find
patterns in the play of others when the game is played repeatedly
\citep[cf.][]{Arthur1994, Meyer_et_al_1992}. The learning model
proposed in the minority game literature provides a way of
formalizing this notion. In this section, we first introduce the
model, and then discuss its assumptions, relating the learning
model to other learning models in the literature.

\subsection{Model}

The stage game is played repeatedly. After each round of play $t$
of the stage game, the players are informed of the aggregate
action $A(t) := \sum_{i=1}^{2k+1} a_i(t)$, where $a_i(t) \in
\{-1,+1\}$ is the action taken by player $i$ in round $t$.
Furthermore, it is assumed that players only retain the sequence
of the last $m$ winning groups $-\text{sign}[A(t)]$, where $m \in
\mathbb{N}$. Hence, in round $t$, players observe the $m$ most
recent outcomes $h_m(t) = (-\text{sign} [A(\tau)])_{\tau \in
\{t-m, t-m+1, \ldots, t-1\}}$.

A \emph{response mode} $s$ assigns to each information set $h_m
\in \mathcal{H}_m = \{(x_k)_{k=1, \ldots, m} | x_k \in \{-1,+1\}
\}$ an action $a \in \{-1,+1\}$. That is, a response mode $s$
prescribes which action $s(h_m(t)) \in \{-1,+1\}$ to take, for a
given history of play $h_m(t)$ at time $t$. There are $2^{2^m}$
different response modes: there are $2^m$ possible signals $h_m$
of length $m$, and for each signal, there are two possible
actions. For memory length $m$, denote the set of all response
modes by $\mathcal{S}^{(m)}$. An important assumption in the
minority game learning model is that each player $i \in
\mathcal{N}$ is endowed with a subset $S_i$ of all possible
response modes, with for each $i \in \mathcal{N}$ the response
modes in $S_i$ drawn uniformly at random from $\mathcal{S}^{(m)}$,
independently across players. Results are then obtained by
averaging over all possible assignments of response modes. This
endowment is fixed for each player, and all players are endowed
with the same number $n_S \geq 2$ of response modes. An example of
such a subset of response modes for $n_S = 4$ and $m=3$ is given
in Table \ref{table:inputoutput}.

\begin{table} \centering
\begin{tabular}{|rrr|rrrr|}
\multicolumn{3}{c} {History} & \multicolumn{4}{c} {Action} \\
\hline & {$h_m$}& &$s_{i,1}$ &$s_{i,2}$ & $s_{i,3}$ & $s_{i,4}$\\
\hline
$-1$ & $-1$ & $-1$ & $+1$ & $-1$ & $-1$ & $+1$\\
$-1$& $-1 $& $+1$ & $-1$ & $-1$ & $+1$ & $-1$\\
$-1$ & $+1$ & $-1$ &$+1 $& $-1$ & $-1 $& $+1$\\
$-1$ &$+1$ & $+1$ &$-1 $& $+1$ & $-1$& $+1$\\
$+1$ &$-1$ & $-1$ & $+1$ & $+1$ &$+1$& $+1$\\
$+1$ &$-1$ &$ +1$ & $-1$ & $-1$ & $+1$& $+1$\\
$+1$ & $+1$ &$ -1$ & $-1$ & $-1$ & $-1$& $+1$\\
$+1$& $+1$& $+1$ & $-1$ & $+1$ & $-1$& $+1$\\
\hline
\end{tabular}
\caption{An example of a subset of response modes with $m=3$ and
$n_S=4$ for some player $i \in \mathcal{N}$.}
\label{table:inputoutput}
\end{table}

When faced with a history $h_m(t)$, an player has to choose which
of his $n_S$ response modes to use in the next round. Each player
$i$ keeps a \emph{virtual score} $p_{i,\ell}(t)$ for each response
mode $s_{i,\ell} \in S_i$ that reflects that response mode's past
performance. The virtual score of each response mode is updated
after each round, regardless of whether the response mode has been
used or not. When a response mode would have correctly predicted
the winning side, its virtual score is increased with the payoffs
it would have earned, otherwise it is decreased with the same
amount. This means that players do \emph{not} take the effect of
their action on the aggregate outcome $A(t)$ into account. In
determining the virtual score of a response mode, players only
consider whether this response mode would have predicted the
actual outcome correctly, neglecting the question whether playing
this response mode would have affected the outcome.
\begin{example} \label{exam:tipped}
Suppose that the payoffs are of the form~\eqref{eq:g_function}.
Then, the updating rule is:
$$
p_{i,\ell}(t+1)=p_{i,\ell}(t)-s_{i,\ell}(h_m(t))\cdot
\frac{A(t)}{2k+1}
$$
where $\ell \in \{1, \dots, n_S\}$. Suppose that in some round
$t$, player $i$ has chosen action $a_i(t) = -1$, and that the
total number of players choosing action $a = -1$ is $k+1$, i.e.,
$-1$ is the majority action. Then the virtual score of all
response modes prescribing $a = -1$ would be decreased by $(k+1)-k
= 1$, while the virtual scores of all other response modes would
be increased by $1$. However, if player $i$ would have played one
of those response modes, the number of players choosing $a = +1$
would have been $k + 1$, and $+1$ would have been the majority
action.
\end{example}
The probability that player $i \in \mathcal{N}$ chooses the
response mode $s_{i,\ell} \in S_i$ in the next round is given by
the well-known logit choice rule:
\begin{equation}
\label{eq:thermal} \text{Prob}\{s_i(t) = s_{i,\ell}\} =
\frac{e^{\beta \cdot p_{i,\ell}(t)}}{\sum_j e^{\beta \cdot
p_{i,j}(t)}}.
\end{equation}
The parameter $\beta$ can be interpreted as the sensitivity of
choice to marginal information. In the limiting case $\beta \to
\infty$, play becomes fully deterministic in the sense that
players choose the response mode with the highest virtual score.
Allowing for $\beta < \infty$ adds noise at the individual level
as well as it introduces additional heterogeneity
\citep{Cavagna_et_al_1999}. When $\beta \to \infty$, all players
endowed with a certain response mode keep the same virtual score
for that response mode. By contrast, for finite $\beta$, players
differ in their ranking of response modes, as their endowment of
response modes determines the denominator of
Equation~\eqref{eq:thermal}. Perhaps surprisingly, this added
heterogeneity and noise actually improves collective performance,
as discussed in Section~\ref{sec:volatility}.

\begin{figure}
\begin{center}
\includegraphics[height=0.45\textwidth]{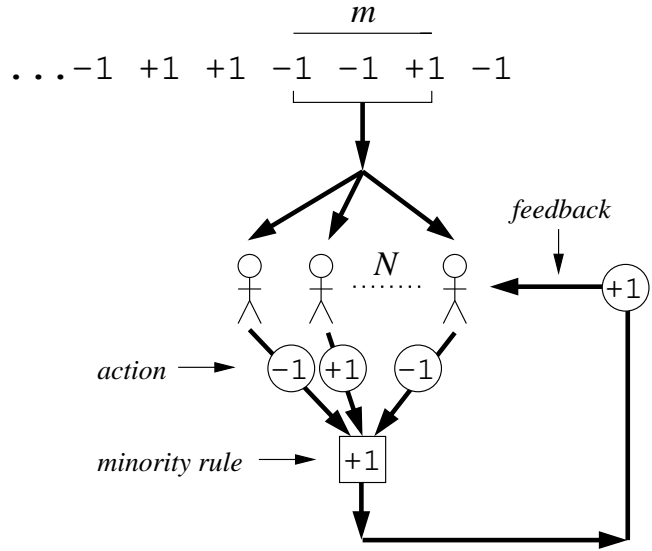}
\vspace*{3mm} \caption{A schematic overview of the minority game
learning model. Figure taken from \citet{Moro2003}. }
\label{fig:feedback_m}
\end{center}
\end{figure}

Actions, outcomes and performance are thus linked by a complex
feedback system, as illustrated in Figure~\ref{fig:feedback_m}.
Players observe the recent outcomes, and choose a response mode
with a probability depending on the number of virtual points that
response mode collected, resulting in an action $a \in \{-1,
+1\}$. The actions of all players determine the winning side
through the minority rule; this information is then fed back to
the players and adds to the sequence of outcomes.

\subsection{Discussion} \label{sec:disc_ass}

In this section, we discuss two of the most important assumptions
of the learning model in the minority game model: the assumption
that all players are endowed with a random subset of response
modes and the assumption that players update the virtual scores of
response modes not used, without taking into account the effect of
that response mode on the game's outcome. Although the learning
model of the minority game literature seems to depart markedly
from the standard evolutionary and learning models used in
economics, we argue here that in fact the learning model combines
different aspects of several game-theoretic models to provide a
realistic model of player behavior in congestion games.

\subsubsection{Response modes and heterogeneity}

In the learning model proposed in the minority game literature,
players base their action on the recent past, trying to discern
patterns in their opponents' behavior, as in \citet{Arthur1994}.
\citeauthor{Arthur1994} proposes that players condition their
decision to go to a bar on attendance levels in the previous
weeks. He employs the terms ``predictor'' or ``hypothesis'' rather
than response mode: if the bar has been crowded for the last three
weeks, I expect it to be crowded next week also. These mental
models are mapped into actions: if I expect the bar to be crowded,
I will not go.

The response modes in the minority game learning model are a
concise way of modelling this notion. An important question,
however, is which response modes need to be included in the model.
There are two possible avenues. Firstly, one could simply
incorporate all possible response modes. However, if all possible
response modes are included in the learning model, the strategy
space becomes huge already for very simple games. Many different
response modes are conceivable in a simple game such as the
minority game, as illustrated by the list of examples
in~\citet{Arthur1994}.

A second possibility is to include only a selection of possible
response modes. In that case, one could either make a selection
based on behavioral assumptions, or let the subset of response
modes be determined randomly. In the first case, a natural choice
is to include response modes that reflect beliefs about other
players' actions, based on recent outcomes. The first approach is
commonly taken in the economics literature
\citep[e.g.][]{ErevRapoport1998,Selten2002}, while the minority
game learning model takes the second avenue. When players'
response modes are drawn uniformly at random from the set of all
possible response modes, there are no restrictions on the types of
response modes that players use.

At first sight, this may seem to be a weak point of the model, as
response modes do not need to have a sensible interpretation in
the minority game learning model. However, it can be shown that
regardless which response modes players are endowed with, players
will self-organize into groups that use different response modes
in such a way that their actions cancel out (see
\citet{ChalletZhang1998, Hartetal2001}; see also
Section~\ref{sec:crowds}). Hence, the minority game learning model
provides a possible explanation for the simultaneous evolution of
behavioral rules (e.g. ``switch roads if the road was crowded in
the previous period'') and their antagonists (``stay at the same
road if the road was crowded in the previous period'') often
observed in congestion game experiments \citep[e.g.][]{Selten2002}
through the structure of the game and players' heterogeneity. The
strong point of the minority game learning model is exactly that
no assumptions regarding response modes are needed. In games such
as the minority game, whether a response mode is reasonable
\emph{only} depends on the response modes used by
others.\footnote{For instance, \citet{Selten2002} reports that
some subjects use a ``direct'' response mode in his experiments on
route-choice games, while other subjects use a ``contrarian''
response mode. Subject who use the former response mode will
switch roads if they experienced congestion in the last period,
while subjects using the contrarian response mode stick with their
choice, as they expect other subjects to switch. The important
point to note is that the direct response mode is only sensible if
there are players who use the contrarian response mode and vice
versa.} Conversely, \emph{any} response mode, whether it has a
sensible interpretation or not, will work if opponents use
response modes that recommend them to take the opposite action
(see Section~\ref{sec:crowds}).

Note that the minority game differs in this respect from games
such as the p-beauty contest \citep[p.
156]{Keynes1936}.\footnote{In the p-beauty contest, players have
to choose a number in a certain interval. Players have to guess
what the average choice will be; the player that picks the number
that is closest to some fraction $\varphi < 1$ of the average
choice will win. Suppose players have to choose a number between 0
and 100, and will win with their choice is closest to $\varphi =
2/3$ of the average choice. Then nobody will choose a number
higher than $\frac{2}{3} \cdot 100$, so nobody should pick a
number higher than $\frac{2}{3} \cdot\frac{2}{3} \cdot 100$, and
so on. When players are rational and there is common knowledge of
rationality, the equilibrium choice is 0.} Both in the p-beauty
contest and the minority game, players base their actions on their
beliefs about other players' actions, who in turn base their
actions on \dots, etcetera. While in the p-beauty contest, this
recursion of actions and beliefs ends at a well-defined limit
point, the Nash equilibrium action, there is no such limit point
in the minority game. This means that there is no action in the
minority game that is optimal a priori, as in the p-beauty
contest: if all think that $a = -1$ will be the minority choice,
then all will choose that action.\footnote{Also see
\citet{CamererFehr2006}. \citeauthor{CamererFehr2006} explain
behavior in congestion games and the p-beauty contest using the
cognitive hierarchy approach \citep{CamererHoChong2004,
StahlWilson1995}.} In such a case, agnosticism on the type of
response modes that players use may well provide a more realistic
model of players' reasoning processes than the more restrictive
assumptions employed in different learning models. This offers an
elegant solution to the dilemma signalled by \citet[][p.
873]{ErevRoth1998} that it is virtually impossible to include all
possible behavioral rules, but that selection of specific rules
bears the risk of ``parameter fitting in a model with an enormous
number of parameters''. In the minority game learning model, no
response mode is ruled out on a priori grounds, while sensible
behavioral rules evolve naturally, as the only criterion for a
behavioral rule to be sensible in the minority game is that there
are other players who follow a ``contrarian'' behavioral rule.

However, this approach raises some questions. Firstly, one may ask
why it is assumed that players are heterogeneous in their
endowment of response modes. Perhaps more importantly, one could
ask why players only consider a fixed number $n_S$ of response
modes. Indeed, individual players have an incentive to increase
the number of response modes they use, as that gives them an
advantage over other players \citep{Marsili_et_al_2000}. However,
these assumptions are not uncommon in game-theoretic models of
learning and bounded rationality.\footnote{For instance, in
cognitive hierarchy models \citep{CamererHo1999, StahlWilson1995},
it is assumed that each player is of some exogenously specified
type; players of different types use different strategies. Another
example is the replicator dynamic \citep[e.g.][]{Weibull1995} in
which players are ``programmed'' to play a given strategy.}
Possible justifications for such assumptions include that each
player has different experiences prior to playing the minority
game and therefore deems different response modes more reasonable
than others \citep[cf.][p.4, and references therein]{Aumann1997,
FudenbergLevine1998}, and that boundedly rational player may
prefer to just consider a subset of response modes that have
worked well in the past, rather than considering all $2^{2^m}$
response modes \citep[cf.][]{EllisonFudenberg1993}.\footnote{The
precise value of $n_S$ is irrelevant. The qualitative behavior of
the model is not affected by the choice of $n_S$, as long as there
is some heterogeneity among players
\citep{ChalletMarsiliZhang2004}.}

\subsubsection{The law of simulated effect and boundedly rational
players}

Which response mode players choose from the set of response modes
they are endowed with, is determined by the virtual score of each
response mode. The learning process proposed in the minority game
literature is closely related to the reinforcement learning model
of \citet{RothErev1995} and \citet{ErevRoth1998}. The main
difference between the basic reinforcement learning model of
\citeauthor{RothErev1995} and the learning model of the minority
game literature lies in the updating of the score of strategies or
response modes not played. In the basic reinforcement learning
model, the scores of these strategies are not updated, while in
the minority game learning model, the scores of all response modes
are updated in every period, as in hypothetical reinforcement
learning or stochastic fictitious play
\citep{FudenbergLevine1998}. The assumption that players also
consider the payoffs to strategies or response modes not played
seems to be reasonable. \citet{CamererHo1999} argue on the basis
of theoretical arguments as well as on the basis of experimental
results that players obey not only the ``law of \emph{actual}
effect'', but also the ``law of \emph{simulated} effect'', meaning
that in reinforcement, not only payoffs from strategies that are
actually used count, but also foregone payoffs from strategies not
played.

However, for players to play according to the act of simulated
effect, they need more information than for standard reinforcement
learning.\footnote{Recall that players only need to know their own
payoff to play according to the standard reinforcement learning
model of \citet{RothErev1995}.} In general, to play according to
fictitious play, players need to know the payoff rule as well as
the actions of their opponents in addition to their own payoff.
Even in a game such as the minority game, where the players only
need to know the aggregate choice of other players (and not their
individual choices), calculating foregone payoffs of strategies
not used may be too hard for players that are boundedly rational.
In the minority game learning model, players' bounded rationality
is reconciled with the law of simulated effect by assuming that
players do not take the effect of their own action on the global
outcome into account. In that way, players can account for
foregone payoffs of response modes not used, without having to do
complicated calculations.

At first sight, one may think that for a large number of players,
it does not matter whether players account for their own impact.
However, due to the minority rule, there remains a systematic bias
in the rewarding of response modes, even if the number of players
goes to infinity. The reason is that the virtual score of a
response mode that is currently played is systematically lower
than that of the response modes that are not used. These latter
response modes get a point if they prescribe the current minority
side, even if they would have tipped the minority to the other
side if they would have been played, so that they would have
guessed wrong in reality (cf. Example~\ref{exam:tipped}). As the
response mode that is actually played does not have this
advantage, the response modes that are not played are
systematically favored and hence results depend on whether players
take the effect of their action on the aggregate outcome into
account \citep{Marsili_et_al_2000, MarsiliChallet2001}.

\medskip

\noindent The minority game learning model thus combines features
from several learning models in the literature on learning in
games. However, the minority game learning model makes distinctly
different predictions than game-theoretical learning models. To
these predictions we now turn.

\section{Predictions of the learning model}\label{sec:results}

In this section, we discuss the main predictions on the minority
game learning model. In the first two sections, we characterize
the behavior of the model in terms of social efficiency and
informational efficiency, and show that the two are inherently
linked in the minority game learning model. In Section
\ref{sec:crowds}, we show how different response modes may evolve,
and discuss the implications for efficiency.

\subsection{Volatility and attendance} \label{sec:volatility}

\begin{figure}[tbp]
\begin{center}
\includegraphics[height=0.42\textwidth]{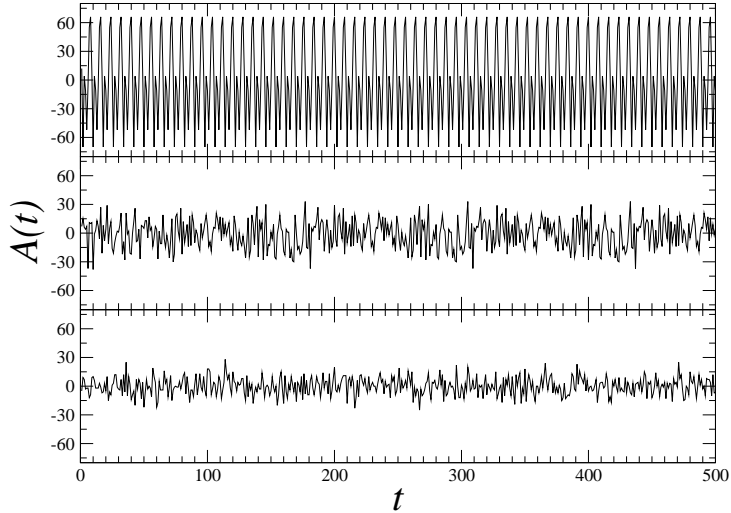}
\end{center}
\caption{Time evolution of the attendance $A(t)$ with $g[A(t)] =
A(t)$, $2k + 1=301$ and $n_S=2$. Panels correspond to $m=2, 7, 15$
from top to bottom. Figure taken from \citet{Moro2003}.}
\label{fig:attendance}
\end{figure}

Typically, dust never settles down in the minority game learning
model: the \emph{aggregate attendance} $A(t):= \sum_{i \in
\mathcal{N}} a_i(t)$ as a function of round number $t$ keeps
fluctuating, as can be seen in Figure \ref{fig:attendance}. As the
game is symmetric, the time average of $A(t)$ will be 0 in the
steady state, as borne out by simulations \citep[see
e.g.][]{ChalletZhang1997, ChalletZhang1998, Johnsonetal1998,
Manucaetal2000}. More interesting is the behavior of the variance
$\sigma^2 := \langle A^2 \rangle$, where $\langle \cdot \rangle$
denotes the (time) average of a quantity. The variance, or
volatility, is a measure of the degree of efficiency achieved in a
population. The higher the variance, the larger the aggregate
welfare loss: large fluctuations around the time average $\langle
A \rangle = 0$ imply that the size of the minority is only small.
When payoffs are linear in $A(t)$, this is easy to see: in that
case, total payoffs are proportional to $-\sum_{i \in \mathcal{N}}
a_i A(t) = -\bigl(A(t)\bigr)^2$.

\begin{figure}[tbp]
\begin{center}
\includegraphics[height=0.3333\textheight]{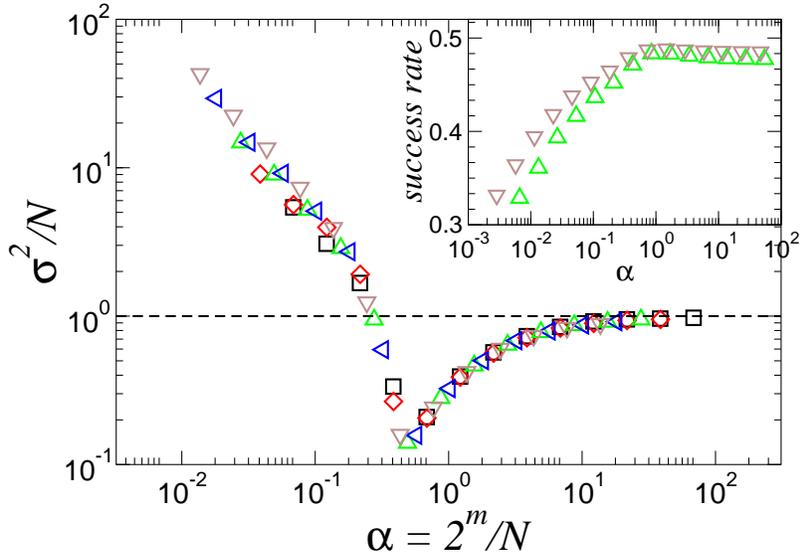}
\end{center}
\caption{Volatility as a function of the order parameter $\alpha$ for $n_S=2$ and different number of players $N:=2k+1=101,201,301,501,701$ ($\square$, $\lozenge$, $\triangle$, $\lhd$, $%
\triangledown$, respectively). The critical value $\alpha_c$ is
the value of $\alpha$ for which the volatility is at a minimum.
Inset: Agent's mean success rate as function of $\alpha$. Figure
taken from \citet{Moro2003}. } \label{fig:original}
\end{figure}

It has been found that $\sigma^2$ is only a function of $\alpha :=
2^m/(2k+1)$ for a given value of $n_S$, where we recall that $n_S$
is the number of response modes of each player
\citep{SavitManucaRiolo1999}. Figure~\ref{fig:original} shows the
volatility as a function of $\alpha$. As can be seen in the
figure, the volatility converges to the volatility exhibited in
the symmetric mixed-strategy Nash equilibrium for $\alpha \to
\infty$. With a large number of players ($\alpha$ small), overall
performance is much worse; in fact, the volatility is of the order
of $(2k+1)^2$, so that the size of the winning group is much
smaller than $k$. At intermediate values of $\alpha$, volatility
is low, and it attains a minimum at $\alpha_c(n_S) \cong n_S/2 -
0.66$ \citep{Marsili_et_al_2000}. Hence, at intermediate values of
$\alpha$, players are able to coordinate their actions and perform
better collectively than under the symmetric mixed-strategy Nash
equilibrium. This means that players can exploit the available
information to predict future market movements so that the
aggregate welfare loss $\sigma^2$ is reduced relative to the
symmetric mixed-strategy Nash equilibrium. Note that this is not
the result of some form of cooperative behavior of the players:
agents are selfishly maximizing their own return, and in doing
that, they come closer to global efficiency.

However, coordination is not complete under the current learning
model. In the socially efficient outcome, players would play
according to one of the pure-strategy Nash equilibria of the game,
and the minority would consist of $k$ players. In that case,
almost half of the players are in the minority, and
$\sigma^2/(2k+1)= 1/(2k+1)$. Players come close to this optimum at
$\alpha = \alpha_c$, although they never reach it. For smaller
values of $\alpha$, performance is much worse than under this
optimum, while for large values of $\alpha$, aggregate payoffs are
close to those of the symmetric mixed-strategy Nash equilibria
(see Figure~\ref{fig:original}). By contrast, when players do take
the effect of their own action on the aggregate outcome into
account, play converges to one of the pure-strategy Nash
equilibria of the game so that coordination is complete
\citep{Challet_et_al_2000, Marsili_et_al_2000, MarsiliChallet2001,
DeMartinoMarsili2001}.\footnote{Also, \citet{KetsVoorneveld2007}
show that most standard learning processes such as the replicator
dynamic converge to the pure-strategy Nash equilibria of the
game.}

Strikingly, global efficiency is enhanced for certain values of
$\alpha$ when players do not always choose the response mode $s$
with the highest number of virtual points, i.e. when $\beta <
\infty$ in Equation~\eqref{eq:thermal}. It can be shown that for
$\alpha < \alpha_c$ (the socially inefficient regime), volatility
\emph{decreases} when the noise level \emph{increases}. For
$\alpha > \alpha_c$, the value of $\beta$ does not affect the
level of volatility \citep{Cavagna_et_al_1999,
ChalletMarsiliZecchina2000, BotazziDevetagDosi2001, Marsili2004}.
This result is not so surprising, however, if one recalls that in
the minority game learning model, rational players herd in the
socially inefficient regime ($\alpha < \alpha_c$). When $\alpha <
\alpha_c$, there are few response modes relative to the number of
players. In that case, players have to crowd at a limited number
of response modes, leading to a large number of players choosing
the same alternative (see Section \ref{sec:crowds}). Setting
$\beta < \infty$ is equivalent to slowing down the updating of
virtual scores for response modes more slowly. A finite $\beta$
therefore acts as a brake against overreaction
\citep{BotazziDevetagDosi2001}.\footnote{This result is
reminiscent of the findings of \citet{Goeree_et_al_2004} who show
that payoff-dependent noise in the decision process is able to
break the cascades that would result otherwise in a social
learning model.}

To summarize, the minority game learning model is characterized by
competition and coordination. Agents compete in trying to exploit
asymmetries in the games outcome, but at the same time, they try
to reduce volatility, as volatility harms all players. Hence,
there is a tension between competition and coordination. These two
are intimately linked in the minority game learning model, as are
information and efficiency. We discuss these issues in more detail
in the next section.

\subsection{Information and efficiency}
\label{sec:information}

As discussed in the previous section, players seem to be able to
coordinate reasonably well for some parameter configurations. The
only way players can interact is through the virtual scores of
their response modes, implying that there is some information in
these values \citep{ChalletZhang1998}. This observation led some
authors to study the information contained in the history of play.
The information content of the history of play, or the degree of
predictability can be measured by \citep{ChalletMarsili1999}
$$
H:=\frac{1}{2^m} \sum_{\nu = 1}^{2^m} \langle A(t+1) \lvert h_m(t)
= \nu \rangle^2,
$$
where the time average of $A(t+1)$ is conditioned to the
requirement that the last $m$ winning groups are given by
$h_m(t)$. If $A(t+1)$ and $h_m(t)$ are independent, then $H=0$.
Loosely speaking, $H$ measures the information in the time series
of $A(t)$. If $H> 0$, then the signal $A(t)$ contains information.
It can be shown that players in the minority game learning model
minimize the degree of predictability
\citep{ChalletMarsiliZhang2004}. Depending on the value of
$\alpha$, they are more or less successful in doing that. At
$\alpha_c$, the system changes from an informationally efficient
and socially inefficient phase ($H = 0$, $\sigma^2$ large) to an
information-rich and socially efficient phase ($H
> 0$, $\sigma^2$ small). In the informationally efficient
phase, players do worse than players playing according to the
symmetric mixed-strategy Nash-equilibrium. By contrast, in the
information rich phase, players manage to coordinate and do better
than players who play according to the symmetric mixed-strategy
Nash equilibrium.

\begin{figure}
\begin{center}
\includegraphics[height=0.45\textwidth,clip=]{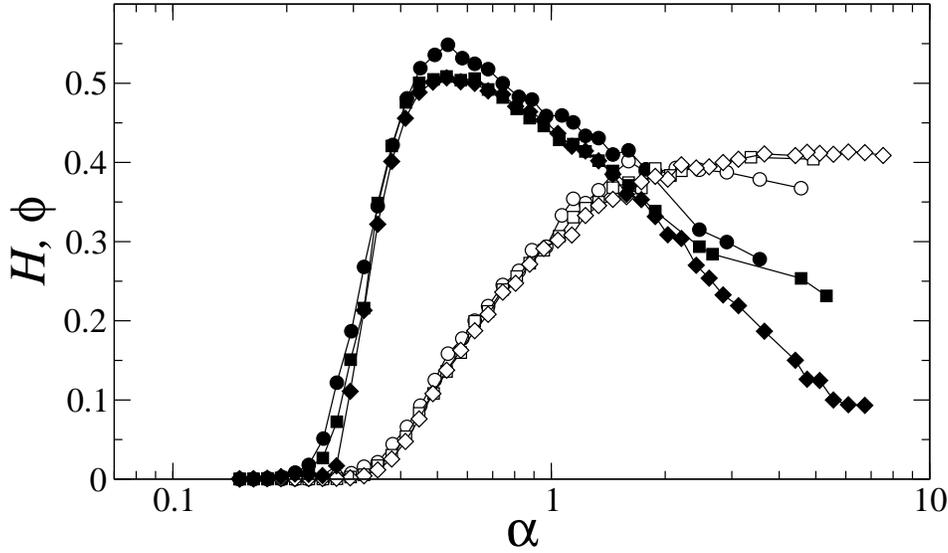}
\caption{Information $H$ (open symbols) and fraction of frozen
players $\phi$ (full symbols) as a function of the control
parameter $\alpha = 2^m/(2k+1)$ for $n_S=2$ and $m = 5,6,7$
(circles, squares and diamonds, respectively). Figure taken from
\citet{Moro2003}.} \label{fig:information}
\end{center}
\end{figure}

At $\alpha = \alpha_c$, the symmetry between the two actions is
broken. In the so-called symmetric phase ($\alpha < \alpha_c$),
both actions are equivalent. Both actions are taken by the players
with equal frequency. For $\alpha > \alpha_c$, one of the actions
is preferred, i.e. the outcome is asymmetric. An asymmetry in the
game's outcome represents an opportunity that could in principle
be exploited. Hence, this is just a concomitant feature of the
presence or absence of information in the history of play.

As an alternative to $H$, one could also consider the fraction of
frozen players \citep{ChalletMarsili1999}. Frozen players are
players who never change their response mode in the stationary
state (in the limit of $\beta \to \infty$ in
Equation~\eqref{eq:thermal}). That is, these players have one
response mode that outperforms all others.\footnote{Note that this
does not imply that these players take the same \emph{action}
always: a response mode is a function of past play, hence the
actions vary with the history $h_m$.} As can be seen from Figure
\ref{fig:information}, the fraction $\phi$ of frozen players is
zero in the informationally efficient phase, while it first rises
for intermediate values of $\alpha$ and then falls again when
$\alpha$ goes to infinity. The intuition is that, for very small
values of $\alpha$ (the informationally efficient phase), both
actions are equivalent, so that there is little variation in the
virtual scores of the different response modes. This means that
players switch response modes easily. For very large values of
$\alpha$, players behave more or less randomly, so they switch
response modes frequently. Only at intermediate values of $\alpha$
the fraction of frozen players is large, as many players have a
response mode that is superior to other response modes. Note that
the success of a response mode depends on the response modes used
by opponents: a response mode \emph{per se} is not superior, it is
the collective of response modes that is successful \citep[see
also][]{SavitManucaRiolo1999}. In particular, it only pays to be
predictable if others are predictable as well.

This transition between the informationally efficient and the
information rich phase, or equivalently between the socially
inefficient and the socially efficient phase, is central to the
minority game learning model. At this transition, there is a
qualitative change in collective behavior, while the principles
behind the behavior of individuals remain unchanged. For all
values of $\alpha$, players in the minority game learning model
try to outsmart each other, but for low values of $\alpha$, they
are on average less successful. In the next section, we discuss
the interpretation of $\alpha$.

\subsection{Response modes and their antagonists}
\label{sec:crowds}

The former sections have shown that the qualitative behavior of
the system depends only on $\alpha = 2^m/(2k+1)$, not on other
variables such as $n_S$. Moreover, for some values of this
parameter, players are much more successful in coordinating
behavior than for other values. What is the feature of the model
underlying this behavior? We address this question in the current
section. The answer to this question points to an intuitive
interpretation of the model's results in terms of response modes
and their antagonists.

The minority rule forces players to differentiate: if all players
choose the same response mode, all will loose. Agents want to be
as far apart in the space of response modes as possible. However,
there are only $2^{2^m}$ possible response modes for $2k+1$
players. Hence, one would expect that players succeed in
differentiating if $2k+1 \ll 2^{2^m}$, while they behave more like
a crowd when $2k+1 > 2^{2^m}$. So, one would expect a qualitative
change at $2k+1 \sim 2^{2^m}$, rather than at $2k+1 \sim 2^m$, as
observed. The reason that the transition occurs at $2k+1 \sim 2^m$
rather than at $2k+1 \sim 2^{2^m}$ is that two response modes
$s,s'$ only give rise to distinctively different behavior if
either they prescribe different actions for every history of play
(i.e., $s$ and $s'$ are anti-correlated) or if their predictions
are uncorrelated \citep{ChalletZhang1998, Johnsonetal1998,
Hartetal2001}. It can be shown that for every response mode $s$,
the number of response modes that are anti-correlated or
uncorrelated with $s$ is given by $2\cdot 2^m/n_S$
\citep{ChalletZhang1998, Hartetal2001}. Hence, $\alpha$ is
proportional to the inverse of this number.

This leads us to an intuitive interpretation of the model's
results in terms of the interplay between different response
modes. Let $s$ be a response mode, and let $\bar{s}$ be the
response mode that is anti-correlated with $s$. Suppose $N_s$
players use the response mode $s$ at a given time step, and
$N_{\bar{s}}$ players use the anti-correlated response mode
$\bar{s}$ at the same time step. If $N_s \approx N_{\bar{s}}$ for
all anti-correlated pairs $(s, \bar{s})$ of response modes, then
the actions of players using these response modes effectively
cancel and the volatility will be small.

Hence, it would be optimal if the group of players that use a
certain response mode is of about the same size as the group that
uses the ``antagonistic'' response mode. However, this is not
always possible, as the dimension of the space of response modes
is fixed by the parameter $m$. Hence, players can only be ``far
apart'' in terms of response modes if the number of players is not
too large relative to the dimension of the response mode space.
For a given number of players, players cannot differentiate if $m$
is small, as the space of response modes is too crowded in that
case. The players display herding behavior: for a pair of
anti-correlated response modes $(s,\bar{s})$, almost all players
herd at one of them, with very few players choosing the other.
Hence, the actions of the players choosing a given response mode
do not cancel those of the players using its antagonist, so that
$\sigma^2$ will be large. For somewhat larger $m$ (for a fixed
number of players), players can differentiate, and the actions of
players effectively cancel. Hence, the system is quite successful
collectively at intermediate values of $\alpha$, although the
minority rule prevents the system from attaining full efficiency,
i.e., not all players can be on the minority side. For a given
$h_m$, the response modes of most players are uncorrelated, but a
small share of players uses response modes that are mutually
anti-correlated. This coordinated avoidance is beneficial for
everybody, as it helps to get a more even division of players over
both alternatives \citep{Zhang1998}.

Now, for very large $m$ at a fixed number of players, the number
of players using a given response mode will only be small, so that
players act more or less independently \citep{Moro2003}. However,
the system still performs better than players who play the
symmetric mixed-strategy Nash equilibrium would, as there always
exists pairs of players that follow anti-correlated response
modes, so that the players' actions are never truly independent
and $\sigma^2/(2k+1)$ is smaller than 1, the value of
$\sigma^2/(2k+1)$ under the symmetric mixed-strategy Nash
equilibrium \citep{ChalletZhang1998}.

\section{Comparison to experimental results} \label{sec:experiments}

In this section, we discuss some experiments on the minority game
and related congestion games. In addition to the minority game, we
focus on market entry games and route-choice games. First, we
briefly introduce these two classes of games. We then present some
experimental results, and discuss whether and how the learning
model proposed in the minority game literature could explain these
results.

\medskip

\noindent The market entry game \citep{SeltenGuth1982} has been
studied extensively in economics.\footnote{See e.g.
\citet{DuffyHopkins2005}, \citet{Kahneman1988},
\citet{Rapoport1995}, \citet{Sundali_et_al_1995},
\citet{ErevRapoport1998}, \citet{Rapoport_et_al_1998,
Rapoport_et_al_2000}.} In a market entry game, $N \in \mathbb{N}$
players must decide independently and simultaneously to enter a
market with a fixed capacity $c < N$ or to stay out. Players who
enter the market receive a payoff that decreases in the number of
entrants. The payoff of players who stay out of the market is
commonly taken to be constant. The game generally has a large
number of Nash equilibria, both in pure and in mixed strategies.
Depending on the exact form of the payoff function, there may even
be a continuum of equilibria. Pure-strategy Nash equilibria may be
payoff-symmetric or payoff-asymmetric, and strict or non-strict,
depending on the choice of parameters. For the payoff functions
commonly studied, the number of entrants is between $c-1$ and $c$
in equilibrium \citep{ErevRapoport1998, DuffyHopkins2005}. An
important difference between the market entry game and the
minority game is that in the latter game, congestion effects are
symmetric, while in the former game, players can choose between a
safe option with guaranteed payoffs -- staying out -- and
entering, the payoffs of which depends on the number of other
players that enter.

As the market entry game is a congestion game, the fictitious play
process converges in beliefs to one of the Nash equilibria of the
game \citep{MondererShapley1996b}. \citet{DuffyHopkins2005} show
that the evolutionary replicator dynamic converge to one of its
rest points, and that the mixed-strategy Nash equilibria of the
game are unstable under the dynamic. They also show that under
standard reinforcement learning \citep{RothErev1995}, the learning
process converges with probability one to one of the pure-strategy
Nash equilibria of the game (when $c \not \in \mathbb{N}$). Under
hypothetical reinforcement learning, where also the propensities
of strategies not used are updated, the learning process converges
with probability one to one of the (logit) perturbed equilibria
corresponding to the pure-strategy Nash equilibria of the game for
$c \not \in \mathbb{N}$ \citep{DuffyHopkins2005}.\footnote{The
perturbed equilibria are the logit quantal response equilibria
\citep{McKelveyPalfrey1995} of the game.}

Route-choice games are closer to the minority game in that there
is no safe option. In a route-choice game, players choose between
two or more roads. The payoffs of choosing one of these roads
falls in the number of other players who have chosen that road.
Roads may differ in terms of capacity. In equilibrium, players
divide themselves over the roads in such a way that traveling
times and hence payoffs are equalized. These games have been
studied experimentally by a number of authors.\footnote{See e.g.
\citet{Iida_et_al_1992}, \citet{HelbingSchoenhofStark2005}, and
\citet{Selten2002}.} An important difference with the minority
game is that the pure-strategy Nash equilibria of the route-choice
game are payoff-symmetric. Moreover, these Nash equilibria are
strict, unlike in the minority game. It is easy to see that the
fictitious play process converges in beliefs to one of the Nash
equilibria of the game \citep{MondererShapley1996b}. No other
analytic results are available on the behavior of different
learning processes in this type of games; however, given the
similarities with market entry games and the minority game, we may
expect learning processes to behave similarly in these games.

The minority game has been discussed in detail in
Section~\ref{sec:game_theoretic}. \citet{KetsVoorneveld2007} study
the predictions of different learning models for the minority
game. They show that the collection Nash equilibria with at most
one mixer is asymptotically stable under the multi-population
replicator dynamic, while other stationary states of the
replicator dynamic are not Lyapunov stable
\citep[e.g.][]{Weibull1995}. Finally, as in all congestion games,
the fictitious play process converges in beliefs to one of the
Nash equilibria of the game.

\medskip

\noindent We now discuss some experimental results on market entry
games, route-choice games and the minority game, and whether, and
how, these results can be explained by the minority game learning
model. A robust finding in experiments on these games is that
subjects quickly achieve a ``magical'' degree of coordination.
However, individual players generally do not play equilibrium
strategies. For instance, while \citet{ErevRapoport1998} find that
the number of entrants in a market entry game rapidly converges to
the equilibrium value, they also observe large between- and
within-subject variability, which does not diminish with
experience. This is a common finding in experiments on market
entry games \citep[][p. 169]{Ochs1999}.\footnote{An exception is
\citet{DuffyHopkins2005} who find that subjects coordinate on one
of the pure Nash equilibria of the market entry game after a large
number of rounds when they are given feedback on others' choices.}
Similarly, in experiments on a route-choice game,
\citet{Selten2002} observe that the mean number of drivers on the
different roads is very close to the equilibrium number, while
large fluctuations persist until the end of the session. Similar
experimental results have been reported for the minority game
\citep{ChmuraPitz2004, PlatkowskiRamsza2003, BottazziDevetag2007}.
In all cases, the hypothesis that fluctuations can be explained by
a symmetric mixed-strategy Nash strategy equilibrium of the game
can be rejected. These results cannot be explained with standard
learning or evolutionary models, as these models typically predict
convergence to the pure-strategy Nash equilibria of such games
\citep{DuffyHopkins2005, KetsVoorneveld2007}. However, as
discussed in Section~\ref{sec:volatility}, the minority game
learning model predicts precisely that average behavior will be
close to the equilibrium prediction, while fluctuations will
persist.

Some authors attempt to reconcile aggregate ``equilibrium''
behavior in experiments with individual non-equilibrium play by
conjecturing that subjects may use counteracting behavioral
rules.\footnote{See \citet{BottazziDevetag2007},
\citet{ChmuraPitz2004}, \citet{ErevRapoport1998},
\citet{Rapoport_et_al_2000}, \citet{Selten2002}, and
\citet{ZwickRapoport2002}.} For instance, \citet{Selten2002}
report that some subjects revise their choice if the road of their
choice turned out to be congested, while other players stick with
their choice in that case, as they expect others to switch. Also
\citet{BottazziDevetag2007} find that there is considerable
heterogeneity in players' behavior in their experiments on the
minority game. They show that it is not the heterogeneity per se
which determines the players' success in coordinating, rather, it
is the interaction between these different behavioral rules that
players can successfully coordinate on choosing different actions.
These findings are in line with the predictions of the minority
game learning model that response modes and their antagonists
coevolve in such a way that their actions effectively cancel out,
thus reconciling aggregate equilibrium behavior and individual
non-equilibrium play.

However, it is not fully clear which behavioral rules subjects
employ. For instance, \citet{Selten2002} are unable to classify
42\% of the subjects in terms of the behavioral rules they use in
their route-choice experiments. This leaves open the possibility
that subjects use some response modes that may not have an
intuitive interpretation and are thus not recognized by the
experimenters, but that nevertheless perform well as response
modes and their antagonists coevolve, as predicted by the minority
game learning model (see Section~\ref{sec:disc_ass} and
\ref{sec:crowds}). A systematic study of the different response
modes used by experimental subjects seems needed. Indeed,
\citet{ZwickRapoport2002} conclude that there is a need ``to
re-orient research on interactive decision making to individual
differences, identify patterns of behavior shared by subsets of
players~\dots, and then attempt to account for aggregate behavior
in terms of the behavior of the clusters of players that form
these aggregates''.

Finally, the effect of information on players' behavior in such
games remains a puzzle. Two dimensions of information have been
investigated in the experimental literature. Firstly, it has been
studied how behavior depends on the information given on other
players' choices. Players can be provided with information only on
the payoff rule and aggregate behavior in the past rounds or may
be informed additionally of the individual choices of all other
players. If players learn e.g. according to the standard
reinforcement learning model of \citet{RothErev1995}, hypothetical
reinforcement \citep{DuffyHopkins2005}, the minority game learning
model, or if the learning process can be described by the
replicator dynamic, this should not affect results.

However, in many experimental studies, behavior differs
qualitatively depending on the information players have.
\citet{DuffyHopkins2005} reports that behavior becomes less random
when players are provided with information on the individual
choices of other players: the hypothesis of randomizing behavior
can be rejected for a larger share of the players, and subjects
seem to display some inertia in their behavior. However, this may
be due to the fact that the additional information given to
players allows them to play complicated repeated-game strategies:
players may signal their commitment to a certain action. While for
the market entry game, such a signalling strategy pays off, this
is not the case in the minority game.\footnote{For instance,
suppose that $k$ players commit to action $a = -1$, and $k$
players commit to action $a = +1$. The remaining player will not
be deterred from choosing either of those actions by the
commitment of other players, nor does the commitment of these
players guarantee them a positive payoff. A repeated-game strategy
that does pay off in the minority game is one in which players
``take turns'': players alternately choose each of the two actions
in such a way that each player is in the minority roughly half of
the time. Indeed, \citet{HelbingSchoenhofStark2005} find some
evidence of such behavior in their experiments on route-choice
games with small groups, but it is unlikely that players will be
able to successfully play according to such a repeated-game
equilibrium when the number of players is large.} Also, one can
imagine that feelings like regret or envy play a larger role in
the market entry game \citep{ErevRapoport1998}. In that sense,
experiments on the minority game provide a cleaner test of
learning theory. Nevertheless, \citet{BottazziDevetag2007} find
that providing players with additional information on their
opponents' play makes that players switch less often between
different actions. In the treatment with full information on
individual players' actions, players tend to stick more often to
their last period's action, especially when this action was the
minority action. Combined with some heterogeneity in players'
beliefs, this inertia and ``reinforcement'' effect partly explains
players' success at coordinating in the minority game. However,
\citeauthor{BottazziDevetag2007} show that inertia, reinforcement,
and heterogeneity alone are not sufficient: players' strategies
also coevolve, or self-organize to improve aggregate payoffs, as
predicted by the minority game learning model.

A second dimension of information that has been studied in the
literature refers to the salience of information on the recent
history of play. \citet{BottazziDevetag2007} provide players with
a string of past outcomes of varying length. When players are
provided with information on play in more rounds than just the
previous one, aggregate efficiency is significantly improved. They
find that providing players with a string of greater length allows
players to correlate their behavior over a longer time period:
when players are provided with the outcome of the previous round,
there is only a significant relation between present and past
choices for the first two time lags, whereas such a relation hold
for up to three time lags when more information is provided.
Notably, in a treatment where players are provided with a string
of intermediate length and the degree of aggregate efficiency is
highest, play is characterized by a substantial lack of
short-range correlations between current and past actions: players
seem to exploit the additional information to improve their
payoffs.

All together, these experimental studies give some support to the
learning model proposed in the minority game literature. However,
the question how information influences play in congestion games
has still not been satisfactorily answered. It would be
interesting to compare players' behavior under different
informational treatments in different congestion games. While most
learning models make similar predictions for the different
congestion games discussed here, intuitively, one would expect
that information will play a different role in these games, as
emotions like envy and regret will be more important in some games
than in others, and also the scope for repeated-game strategies
differs across games. Such a systematic comparison would allow one
to better separate the learning effects from possible
repeated-game and behavioral effects.

\section{Conclusions}
\label{sec: conclusions}

In this paper, we have given a critical account of the learning
model proposed in the learning model proposed in the minority game
literature, and related it to standard learning and evolutionary
models in economics, showing that it shares quite a few features
with these models. Still, the predictions of this learning model
are markedly different from the predictions from other models.
However, these predictions are in line with some experimental
results on the minority game and related games, which cannot be
explained by other models.

However, our understanding of learning in such games is far from
complete. For instance, the effect of feedback on play is unclear.
An interesting direction for further research would be to
systematically vary players' information in experiments on
different congestion games such as the minority game and the
market entry game, and to compare play under the different
information treatments and across games. While most learning
models provide similar predictions for these games, intuitively,
one would expect that information may have different effect in
these games, as in some games, repeated-game strategies or
emotions may play a larger role than in others. Such an experiment
may help shed light on the question which learning model is
appropriate in such games.

\bibliographystyle{chicago}
\bibliography{bibl_MG}
\end{document}